\documentclass[dvips,twoside,fleqn]{article}
\usepackage{epsfig,espcrc2}

\newcommand{\ot}{\frac{1}{2}}
\newcommand{\SH}{S\!H}
\newcommand{\PI}{P\!I}

\title{
\vspace{-5.0cm}
\begin{flushright}
{\normalsize UNIGRAZ-}\\
\vspace{-0.3cm}
{\normalsize UTP-}\\
\vspace{-0.3cm}
{\normalsize 07-09-95}\\
{\normalsize hep-lat/9509009}\\
\end{flushright}
\vspace*{2.5cm}
Spin and Gauge Systems  on Spherical
Lattices\thanks{Supported by BMBF; the simulations have been
performed in part on Cray-YMP of HLRZ J{\"u}lich.}}

\author{
Ch.~Hoelbling\address{Institut f{\"u}r Theoretische Physik,
Karl-Franzens-Universit\"at Graz, Austria},
A.~Jakovac$^{\rm a}$,
J.~Jers{\'a}k\address{Institut f{\"u}r Theoretische Physik E, RWTH Aachen,
Germany},
C.~B.~Lang$^{\rm a}$\thanks{Speaker at the conference} and
T.~Neuhaus\address{FB8 Physik, BUGH Wuppertal, Germany}}

\begin{document}

\begin{abstract}

We present results for 2D and 4D systems on lattices with topology
homotopic to the surface of a (hyper) sphere $S^2$ or $S^4$.  Finite
size scaling is studied in situations with phase transitions of first
and second order type.  The Ising and Potts models exhibit the expected
behaviour; for the 4D pure gauge $U(1)$ theory we find consistent
scaling indicative of a second order phase transition with critical
exponent $\nu\simeq 0.36(1)$.

\end{abstract}

\maketitle

\section{INTRODUCTION}

Some time ago it was suggested to study certain systems on lattices
with trivial first homotopy group \cite{JeLaNe}.  The D-dimensional
boundaries of (D+1)-dimensional hypercubes have this property, being
homotopic to $S^D$.  Here we give a brief summary of recent results
obtained by various subgroups of the present authors for such studies
for compact $U(1)$ pure gauge theory with mixed action \cite{JeLaNe95},
for the 2D Ising model \cite{HoLa95} and for the 2D Potts model
\cite{JaLaNe95}.

The original motivation was to check the possible relevance of periodic
boundary conditions for the dynamics of looplike excitations (like
monopole loops in compact U(1) gauge theory). In the thermodynamic
limit at the critical point the manifold becomes flat and should lead
to a universal field theory independent of the details of the boundary
conditions.  Other 2D studies on similar lattices include random as
well as tetrahedral lattices \cite{JaKaVi94,DiGoSa94} and a recent
investigation of the 7-vertex model (the strong coupling Schwinger
model) for a pillow-like topology \cite{GaLa95}.

We studied the following lattices:
\begin{description}
\item[${T[L]}$~] periodic b.c., i.e. torus topology for the $L^D$ 
hypercubic lattice
\item[${\SH[L]}$~] the surface of a $D+1$-dimensional $L^{D+1}$ hypercube, or
its dual lattice $\SH'$ (where points of $\SH$ are identified with
$(D-1)$--dimensional objects of $\SH'$ etc.)
\item[${S[L]}$~] like $\SH$ but with weight factors correcting for a
spherelike, smooth
distribution of the curvature over the lattice
\item[${\PI[L]}$~] a pillow-like shape, like the surface of a $L^D\times 1$
hypercube; may be visualized as two copies of a $L^D$ hypercube glued
together at their boundaries
\end{description}
In the finite size scaling (FSS) analyses we use $V^{1/D}$ as the
generic lattice size parameter, where the volume $V$ denotes the
number of sites or plaquettes on 2D or 4D lattices,
respectively.

\section{MONTE CARLO SIMULATIONS}

The Monte Carlo updating details depend on the studied system and are
discussed below. In all cases bulk quantities were obtained at various
values of the coupling and combined with the multihistogram method
\cite{FeSw}, giving
\begin{equation}
Z_L(\beta) = \sum_E \rho_L(E)\exp{(-\beta E)}\;.
\end{equation}
This allows us to obtain continuous curves for the observables and to
determine the complex positions of the partition function zeroes.
Also the combined data for the ``multihistogram'' provides for
a reliable identification of possible 2-state signals in the
distribution.

We studied the peak positions (pseudocritical couplings $\beta_{c,V}$)
and peak values of the cumulants
\begin{eqnarray}\label{cumdefs}
C_V(\beta,L)&=& ~~\frac{1}{V} \langle (E-\langle E\rangle )^2
\rangle \; ,
\\
V_{BCL}(\beta,L)&=& -\frac{1}{3} \frac{\langle (E^2-\langle
E^2\rangle )^2\rangle}{\langle E^2\rangle^2} \; ,\\
U_4(\beta,L)&=& \frac{\langle (E-\langle E\rangle)^4\rangle}
{\langle(E-\langle E\rangle )^2\rangle^2}  \; ,
\end{eqnarray}
and the positions $z_0$ of  the Fisher zeroes closest to the real axis.

For 1$^{st}$ order transitions one expects the FSS  behaviour
\begin{eqnarray}
{C_{V,max}\over V}&\rightarrow&  ~~\frac{1}{4}(e_o-e_d)^2 \; ,\\
V_{{BCL},min}  &
\rightarrow& -\frac{1}{12}{(e_o^2-e_d^2)^2\over (e_oe_d)^2} \; ,\\
U_{4,min}  &\rightarrow& 1\;  \\
\beta_{c,V}  &=&\beta_c + a  V^{-1} + O(V^{-2/D}) \label{pscfss}
\end{eqnarray}
with ``surface'' corrections of $O(V^{-2/D})$ due to the
inhomogeneities of the boundary (curvature); in $D=4$ this is
the leading term \cite{JeLaNe}.
The parameter $a$ in (\ref{pscfss}) depends
on the considered cumulant.

For 2$^{nd}$ order transitions we expect
\begin{eqnarray}
C_{V,min} &\simeq &V^{\alpha/D\nu} \\
V_{BCL,max} &\simeq &V^{\alpha/D\nu - 1}  \\
\beta_{c,V}  &=&\beta_c + a V^{-1/{D\nu}} + O(V^{-2/D}) \\
Im \;z_0(V) & \simeq& V^{-1/D\nu}
\end{eqnarray}
with Josephson's law $\alpha = 2 - D\nu$
(for Gaussian systems $\nu = \ot, \alpha = 0$; for a discontinuous
transition $\nu =1/D$). Again in $\beta_{c,V}$ one expects
``surface'' corrections of $O(V^{-2/D})$.

\section{2D ISING MODEL}

This was studied in \cite{HoLa95} in order to identify the FSS effects
of the new topology in more detail. We used lattices of various size
and of type $T$ (where there exists an analytic solution),
$\SH'(16)\ldots \SH'(128)$, $S(16)\ldots S(128)$ and $\PI(16)\ldots
\PI(128)$. The updating was with the Swendsen-Wang Cluster algorithm
with typically $10^7$ measured configurations (with an autocorrelation
time $< 5 $) per lattice size.

The results were typical for the other systems as well. Since $\nu=1$
one expects a leading FSS behaviour $O(1/L)$.  The pseudocritical
points for the spherelike topologies are -- compared to the torus
values -- very close to their thermodynamic values and exhibited a {\em
very small} contribution $O(1/L)$; the scaling behaviour is dominated
by the non-leading term $O(1/L^2)$ which we interpret as coming from
the boundary or curvature, respectively. This holds also for the real
part of the closest Fisher zero.

\begin{figure}[htb]
\vspace{-10mm}
\begin{center}
\epsfig{file=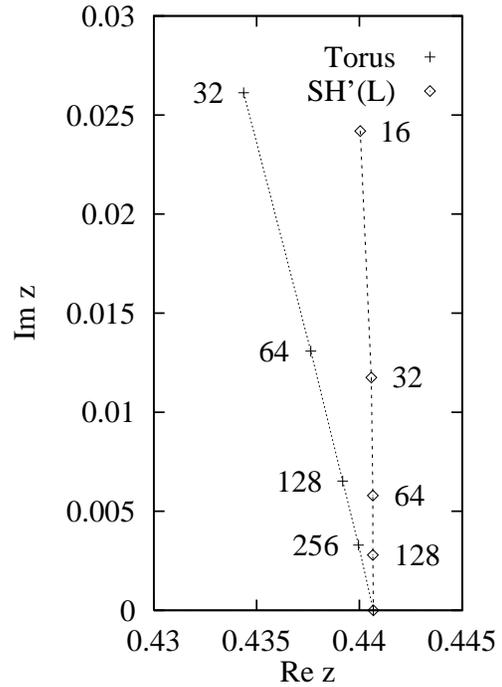,width=68mm}
\end{center}
\vspace{-8mm}
\caption{Ising model: closest Fisher zeroes.}
\label{fig1}
\end{figure}
\vspace{-4mm}

Both, the peak value of $C_V$ as well as $Im z_0$, however, show
scaling in perfect agreement with $\nu=1$. From fig. \ref{fig1} we
find, that for the spherelike topology the zeroes approach the real
axis in the complex coupling constant plane almost perpendicularly.
Thus the real part (related to the pseudocritical points derived from
the cumulants) is always very close to the thermodynamic value with its
scaling behaviour dominated by the $O(1/L^2)$ term.

\section{2D POTTS MODEL}

In \cite{JaLaNe95} the $q$-state Potts model for $q=5, 7, 10$ at the
(for $q=5$ extremely weak) 1$^{st}$ order transition is studied with
multicanonical updating \cite{BeNe}, typically $10^7$ updates for each
lattice size.  The lattices were of type $T(20)\ldots T(50)$,
$\SH(10)\ldots \SH(22)$ and $\PI(16)\ldots \PI(36)$.

\begin{figure}[htb]
\vspace{-6mm}
\epsfig{file=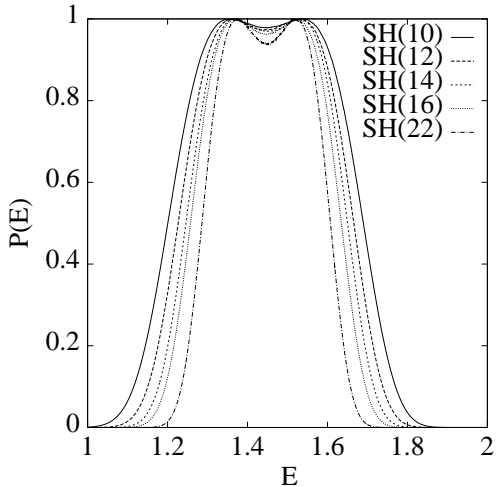,width=68mm}
\vspace{-8mm}
\caption{Potts model (q=5): distribution at equal height.}
\label{fig2}
\end{figure}
\vspace{-2mm}

Even for the weak signal of the $q=5$ model the histogram for the
energy distribution exhibits a clear double peak structure (cf. fig.
\ref{fig2}).  Thus the spherical lattices do no ``wash out'' weak
2-state signals, as might be suspected. Note that there is only one
interface as compared to two for the periodic boundary conditions. This
difference may be of some technical advantage in the determination of
the interface tension.

\section{4D GAUGE THEORY}

The study \cite{JeLaNe} of compact $U(1)$ pure gauge action in the
Wilson formulation with an additional adjoint term (coupling $\gamma$)
has been continued\cite{JeLaNe95}.

For $\gamma=0$, studies for torus geometry find consistently stable
2-state signals, reconfirmed by the recent results of \cite{Wup} and
\cite{KeReWe} (who in particular emphasized the r{\^o}le of monopole
clusters for the phase transition).  Already in 1985 a careful study of
this action at positive $\gamma$ hinted at a tricritical point (TCP)
\cite{TCP}  near $\gamma_{TCP}=-0.11$.  However, 2-state signals have
been observed even below that value \cite{TCP,JeLaNe}.  On the other
hand, studies for spherelike topology show no 2-state signals in the
energy distribution at $\gamma\leq 0$ \cite{JeLaNe} for the studied
lattices $L\leq 12$ (although they do show 2 states at e.g.
$\gamma=0.2$).

Independent of the issue of a possible 2-state signal for the Wilson
action there are no generally agreed results and values for {\em
critical indices} in the $U(1)$ gauge model.

In order to respond to a possible criticism of our original
investigation of the model on $\SH$-lattices -- where the curvature was
concentrated on the edges and corners of the hypercubic lattice -- we
have continued work on an ``almost smooth sphere'' $S(L)$. There we
project sites, links and plaquettes of $\SH(L)$ or $\SH'(L)$ (dual) onto
the surface of 4D spheres and introduce weight factors similar to those
used by \cite{ChFrLe82} in their study of random triangulated lattices,
\begin{equation}
S= - \sum_P\; {A_P' \over A_P} \; [ \beta \cos(U_P) +
\gamma \cos(2 U_P)]\;.
\end{equation}
Here, $A_P$ and $A_P'$  denote the areas of the corresponding plaquette
(and its dual, respectively) of the projected lattice.

More technical details of the simulation have been discussed
\cite{JeLaNe} and will be discussed elsewhere \cite{JeLaNe95}. We have
worked with lattices $S(L)$ for $L$ ranging between 4 and 12.
The couplings were chosen in the immediate
neighbourhood of the critical values of $\beta$ for $\gamma = 0, -0.2,
-0.5$. Since we never observed 2-state signals we did not implement
multi-canonical updating.  For each lattice size we typically
accumulated $10^6$ updates (3-hit Metropolis, for $\gamma=0$ with an
additional overrelaxation step).  Close to the phase transition we
found quite large autocorrelation lengths (300-2000).

We find no indication of 2-state signals in the multi-histogram
distributions.  The values of the cumulants at their respective extrema
are compatible with a continuous phase transition.  Like for the Ising
model, also here we observe in general smaller FS corrections than for
the torus geometry (periodic b.c.). The leading behaviour seems to be
dominated again by the boundary or curvature term, i.e.  suppressed by
$O(1/L^2)$, which at first sight cannot be distinguished from a leading
scaling behaviour for $\nu=0.5$.

The peak values of $C_V$ show power law behaviour and in the table we
give the range of values for $\alpha/\nu$ (and the resulting $\nu$, if
we use Josephson's law) from fits to data for all lattice sizes or
without the largest and/or smallest lattices.  These values appear to
depend on $\gamma$. One has to keep in mind, however, that the FSS law
used for this fit is asymptotic and that there is a regular
contribution to the value of $C_V$ which may be non-negligible for the
lattice sizes entering the fit.

\vspace{-4mm}
\begin{table}[hbt]
\caption{Fit results for $\nu$ from $C_V$ and $Im\;z_0$
for various $\gamma$}
\label{tabfit}
\begin{center}
\begin{tabular}{rccc}
\hline
& \multicolumn{2}{c}{from $C_{V,max}$}&from $Im\;z_0$\\
$\gamma$ &  \multicolumn{1}{c}{$\alpha/\nu$}
& \multicolumn{1}{c}{$\rightarrow \nu $} & \multicolumn{1}{c}{$\nu$} \\
\hline
0.   & 1.50-1.54   & 0.35-0.37& 0.35(1)\\
-0.2 & 1.10-1.38 & 0.37-0.39& 0.37(1)\\
-0.5 & 0.86-1.02 & 0.40-0.41& 0.36(1)\\
\hline
\end{tabular}
\end{center}
\end{table}
\vspace{-8mm}

\begin{figure}[htb]
\epsfig{file=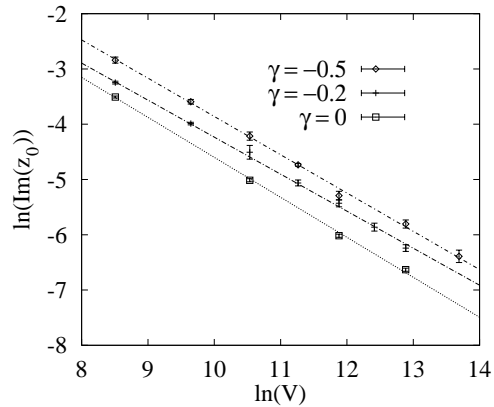,width=68mm}
\vspace{-8mm}
\caption{We find consistent FSS for $Im(z_0)$ for
various values of $\gamma$}
\label{fig3}
\end{figure}

If we study the closest Fisher zeroes (in the complex $\beta$-plane),
we find a behaviour reminiscent to the discussed Ising results.  Fig.
\ref{fig3} exhibits FSS consistent with $\nu = 0.36(1)$ for all three
values of $\gamma$ (cf. the results of the fit in table \ref{tabfit}).
Once again $Im\;z_0$ gives the cleanest and most consistent result.

We should mention, that this value of $\nu$ is compatible with neither
the Gaussian value 0.5 nor the ``discontinuity'' value 0.25; it is
compatible with results obtained in the mid-80s with MCRG methods.
More detailed analyses with improved statistics will be presented
elsewhere \cite{JeLaNe95}.

\end{document}